\documentclass{midl} % Include author names
\usepackage{mwe} % to get dummy images
\usepackage{caption}
\usepackage{graphicx}
\usepackage{booktabs}% to put image and table in the same row

\jmlrproceedings{MIDL 2020}{Medical Imaging with Deep Learning 2020}
\jmlrvolume{}
\jmlryear{}
\jmlrworkshop{MIDL 2020 -- Short Paper}
\editors{}

\title{Joint Liver Lesion Segmentation and Classification via Transfer Learning}

\midlauthor{\Name{Michal Heker\nametag{$^{1}$}} \Email{michalheker@gmail.com}\\
\Name{Hayit Greenspan\midlotherjointauthor\nametag{$^{1}$}} \Email{hayit@eng.tau.ac.il}\\
\addr $^{1}$  Department of Biomedical Engineering, Tel-Aviv University, Israel \\
}

\begin{document}
\maketitle

\begin{abstract}

Transfer learning and joint learning approaches are extensively used to improve the performance of Convolutional Neural Networks (CNNs). In medical imaging applications in which the target dataset is typically very small, transfer learning improves feature learning while joint learning has shown effectiveness in improving the network's generalization and robustness. In this work, we study the combination of these two approaches for the problem of liver lesion segmentation and classification.
For this purpose, 332 abdominal CT slices containing lesion segmentation and classification of three lesion types are evaluated. For feature learning, the dataset of MICCAI 2017 Liver Tumor Segmentation (LiTS) Challenge is used.
Joint learning shows improvement in both segmentation and classification results.
We show that a simple joint framework outperforms the commonly used multi-task architecture (Y-Net), achieving an improvement of 10\% in classification accuracy, compared to a 3\% improvement with Y-Net.

\end{abstract}


\section{Introduction}

% Deep Learning in Medical imaging
Deep learning methodologies, especially Convolutional Neural Networks (CNNs), are the top performers in most medical image processing tasks, including liver lesion  segmentation and classification in abdominal CT images \cite{litjens2017survey},\cite{ben2016fully},\cite{heker2019hierarchical}.
Automatic segmentation and classification is challenging due to different liver lesions contrast, and considerable visual appearance variability between lesion types.
%different types of lesions in terms of lesion shape, texture, and size.

Lesion segmentation has attracted  attention in recent years, with publicly available datasets that enable comparison between different methods \cite{bilic2019liver},\cite{soler20103d}. Lesion classification, on the other hand, is far less investigated with very limited-size datasets explored and no public data available.
% motivation 
%The lack of sufficient amounts of annotated data is one of the main challenges in the medical imaging domain. 
%Another challenge
%that needs to be addressed 
%is class imbalance, which is a key property of multi-class liver lesions.
% What we propose
Transfer learning and joint training are two of the approaches used to address the limited data challenge and improve the performance of CNNs. 
%Transfer learning and fine-tuning have become integral to medical applications, where labeled data is scarce, allowing better feature learning. 
Joint learning, including multi-task learning, has been shown to improve network generalization, resulting in better performance on a given target task \cite{ruder2017overview}.

% Previous work- y-Net, Mammography, pretraining on imageNet

In this work, we compare different approaches to transfer learning with the goal of improving joint segmentation and classification of liver lesions. We introduce and compare the results for two U-Net based frameworks that incorporate joint learning.
% Contributions- transfer learning + joint learning
The following contributions are included in this research:
(1) We focus on transfer learning with the use of data from a similar domain and related target task.
(2) Two U-Net based frameworks that combine transfer learning and joint learning of segmentation and classification are introduced and evaluated. 
% (1) We address the lack of training data by combining joint learning and fine-tuning based transfer learning. 
% (2)Two frameworks for joint learning of segmentation and classification are evaluated and compared to baseline models.
% (3) We emphasize the importance of transfer learning, along with the clear advantage of using data from a similar domain and with a related target task.

%\section{Methods}

%\subsection{Dataset}
% We used two datasets in this work: \textbf{Sheba dataset:} The main dataset is a small in-house dataset, including 332 2D CT slices from the Sheba Medical Center (Israel) with medical records from 140 patients for cases of cysts, metastases, and hemangiomas. \textbf{LiTS dataset}- public dataset, consisting of 130 3D abdominal CT scans ($~$60,000 2D slices) from several different clinical sites with liver and lesion mask annotations (ground truth).

% figure of the system architecture
\begin{figure}[h]%%tbp]
 % Caption and label go in the first argument and the figure contents
 % go in the second argument
\floatconts
  {fig:frameworks2}
  {\caption{The proposed frameworks: (1) SE-Resnet for classification. (2) U-Net with SE-Resnet backbone for segmentation. (3) Y-Net like architecture, i.e. U-Net with SE-Resnet backbone with 2 separate output branches; a classification output at the end of the encoder and a segmentation output at the end of the decoder. (4) U-Net with SE-Resnet backbone for segmentation by pixel-wise classification.}}
  {\includegraphics[width=1\linewidth]{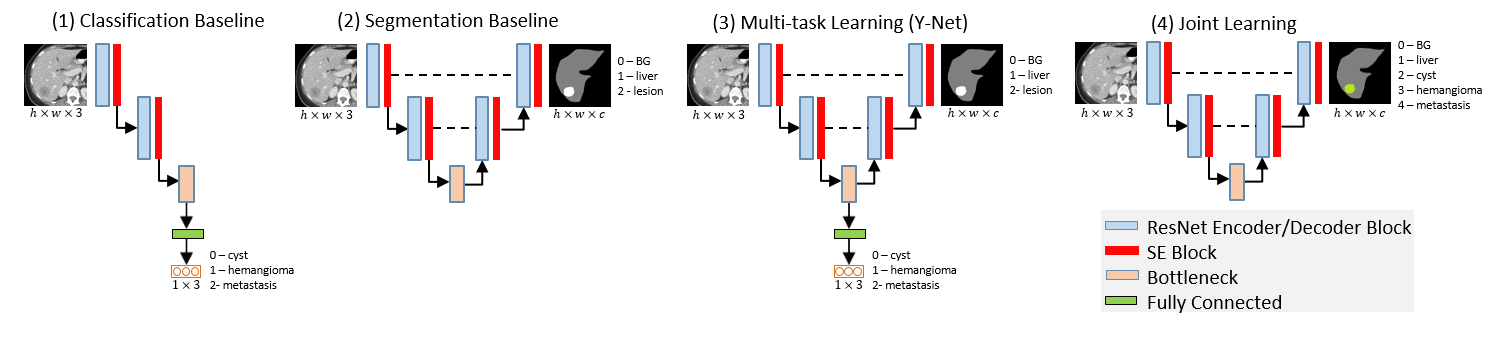}}
\end{figure}

\section{Proposed Frameworks for Joint Segmentation and Classification}
% Explain the frameworks- FCN (U-Net), ResNet backbone, SE Blocks
The proposed frameworks are based on a U-Net model with a SE-ResNet encoder, combining residual skip connections and Squeeze-and-Excitation blocks (SE blocks) \cite{ronneberger2015u}, \cite{he2016deep}, \cite{hu2018squeeze}. The U-Net architecture enables the encoder and the decoder to share information and learn both global and local features to produce quality segmentation.
The first framework incorporates a multi-task U-Net, similarly to \cite{mehta2018net},\cite{le2019multitask}, where joint semantic segmentation and global classification are separated into two distinct outputs and trained simultaneously. The segmentation output is extracted from the U-net's decoder as in a vanilla U-Net, while the classification output is extracted as a parallel branch from the U-Net's encoder. We use a multi-class loss $ L = \lambda{}L_{seg}+(1-\lambda{})L_{cls}$, where weighted cross-entropy loss is used for segmentation to balance the classes (the lesion class is under-represented compared to the liver and background classes), and categorical cross-entropy is used for classification.
% combined loss
In the second framework, segmentation and classification are combined into one output that performs semantic segmentation (pixel-wise classification), where the final lesion class is determined using a majority vote. %pixel-wise prediction map as output % cross-entropy loss
Weighted cross-entropy loss is used here as well, with suitable weights for each class, reversely proportional to their ratio in the dataset.
Additionally, classification and segmentation models were trained individually to get baseline performances on each task separately. The different frameworks are illustrated in \figureref{fig:frameworks2}.

We train and evaluate the models on 332 2D CT slices from Sheba Medical Center (Israel) 
%a private clinical center
%the *anonymous %Sheba Medical Center (Israel) 
with medical records from 140 patients for cases of 3 types of liver lesions with the following distribution: 75 cysts, 71 hemangiomas, 93 metastases. Since the dataset size is small, we use 3-fold cross-validation to evaluate the performance of the models.

%\subsection{Transfer Learning}
We explore the benefit of using transfer learning  by fine-tuning the frameworks in different approaches: %We compare different fine-tuning based transfer learning approaches by training each of the models 3 times: 
(1) training from scratch (random initialization), (2) fine-tuning ImageNet weights \cite{russakovsky2015imagenet}, (3) fine-tuning the weights of a self-trained lesion segmentation model, trained on the dataset of MICCAI 2017 Liver Tumor Segmentation (LiTS) Challenge. The LiTS dataset contains thousands of lesion images, which enables training a strong and robust lesion segmentation model \cite{bilic2019liver}. %The segmentation baseline model is used for training a lesion segmentation network.
% LiTS dataset, which shares a similar domain with our data.

It is worth mentioning that we use liver crops as input during training and not the entire CT slice, which helps the class-imbalance issue. This is done by using a 2-stage cascaded approach, where a first network is trained on the LiTS dataset to obtain high quality liver segmentation (achieving competitive results of 96.1\% in Dice per case score on the LiTS challenge leader board). This network is used on our private dataset to extract liver ROI crops that are used as inputs for the trained models. 

% transfer learning- weights

% cascade

% Loss functions of each system- combined for y-net

% post processing? augmentations? implementation details

\section{Results and Conclusions}
\tableref{table:1} reports the segmentation and classification performance of the frameworks trained with different weights initialization. % on the .%Sheba dataset.
Dice coefficient, Recall, and classification accuracy are used for  evaluation. Qualitative results are presented in \figureref{fig:qualit2}.

% % \begin{document}
%   \begin{minipage}{\textwidth}
%   \begin{minipage}[b]{0.49\textwidth}
%     \centering
%     % \rule{6.4cm}{3.6cm}
%     \includegraphics[width=0.65\linewidth]{qualit2.png}
%     \captionof{figure}{A table beside a figure}
%   \end{minipage}
%   \hfill
%   \begin{minipage}[b]{0.5\textwidth}
%     \centering
%     \scalebox{0.6}{
%         \begin{tabular}{l | l c  c c}
%         \hline
%         Training strategy & Fine-tuning & Cls Acc & Seg Dice  & Seg Recall\\
%         \hline
%          & \textbf{scratch} & 0.55 & - & - \\
%         (1) Cls baseline & \textbf{ImageNet} & 0.63 & - & -  \\ 
%          & \textbf{LiTS}& 0.76 & - & - \\
%         \hline
%          & \textbf{scratch}& - & 0.59 & 0.59 \\
%         (2) Seg baseline & \textbf{ImageNet}& -  & 0.63 & 0.67 \\
%          & \textbf{LiTS}& - & 0.71 & 0.72 \\
%          \hline
%          & \textbf{scratch}& 0.43 & 0.49 & 0.43 \\
%         (3) Y-Net & \textbf{ImageNet}& 0.68  & 0.67 & 0.65 \\
%          & \textbf{LiTS}& 0.79 & 0.71 & 0.68 \\
%          \hline
%          & \textbf{scratch}& 0.63 & 0.57 & 0.60 \\
%         (4) Joint learning & \textbf{ImageNet}& 0.74 & 0.64 & 0.70  \\
%          & \textbf{LiTS}& \textbf{0.86} & \textbf{0.71} & \textbf{0.76} \\
%         \hline
%         \end{tabular}
%         }
%       \captionof{table}{A table beside a figure}
%     \end{minipage}
%   \end{minipage}
% % \end{document}

\begin{table}[h]
\centering
\tabcolsep 25pt % Expand table
\caption{Performance comparison of segmentation and classification}
\scalebox{0.7}{
\begin{tabular}{l | l c  c c}
\hline
Training strategy & Fine-tuning & Cls Acc & Seg Dice  & Seg Recall\\
\hline
 & \textbf{scratch} & 0.55 & - & - \\
1. Classification baseline & \textbf{ImageNet} & 0.63 & - & -  \\ 
 & \textbf{LiTS}& 0.76 & - & - \\
\hline
 & \textbf{scratch}& - & 0.59 & 0.59 \\
2. Segmentation baseline & \textbf{ImageNet}& -  & 0.63 & 0.67 \\
 & \textbf{LiTS}& - & 0.71 & 0.72 \\
 \hline
 & \textbf{scratch}& 0.43 & 0.49 & 0.43 \\
3. Multi-task learning (Y-Net) & \textbf{ImageNet}& 0.68  & 0.67 & 0.65 \\
 & \textbf{LiTS}& 0.79 & 0.71 & 0.68 \\
 \hline
 & \textbf{scratch}& 0.63 & 0.57 & 0.60 \\
4. Joint learning & \textbf{ImageNet}& 0.74 & 0.64 & 0.70  \\
 & \textbf{LiTS}& \textbf{0.86} & \textbf{0.71} & \textbf{0.76} \\
\hline
\end{tabular}
}\label{table:1}
\end{table}

% confusion matrix

% qualitative results

%\section{Conclusion}
% \vspace{-0.1 pt}
\begin{figure}[h]%%tbp]
 % Caption and label go in the first argument and the figure contents
 % go in the second argument
\floatconts
  {fig:qualit2}
 %\vspace{-0.1}
  {\caption{ Lesion segmentation and classification results: (a) input image; (b) ground truth; (c) joint learning; (d) multi-task Y-Net.}}
  {\includegraphics[width=0.4\linewidth]{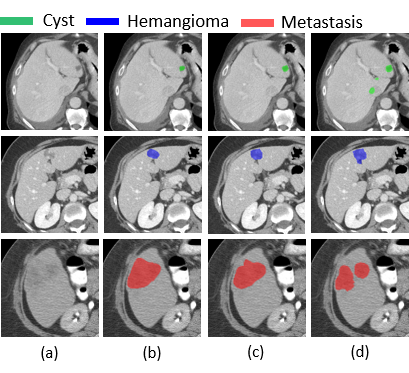}}
\end{figure}

We observe that the joint learning frameworks (see \tableref{table:1}: 3,4) outperform individual task learning.
Interestingly, the simple joint framework achieved the best results in both segmentation and classification tasks, outperforming the commonly used multi-task architecture (Y-Net).
Compared to the baseline models, Y-net showed  3\% improvement in classification accuracy. The joint learning framework showed large improvements: 10\% in classification and 4\% in segmentation recall with the same dice score.
 
%In the Y-net model, features that represent the lesion type class do not share local information with the lesion class, while in the joint model segmentation information is shared with class information.

Using transfer learning generally improved weight initialization and resulted in faster convergence.
ImageNet pre-training showed improved convergence and accuracy.
Pre-training with LiTS enabled the model to learn shared representations in similar domains, resulting in better generalization and higher accuracy. 
%than the ImageNet weights. 

From the evaluation study conducted, we conclude that in the joint network classification and localization context are shared for mutual benefit, thus increasing results. A second conclusion is that transferring learned models via pre-training of networks between two similar tasks provides for stronger and robust representations. In future work, we hope to show the generalization to other domains and tasks.

%The joint learning model outperformed the multi-task framework (Y-Net)

% Acknowledgments---Will not appear in anonymized version
\midlacknowledgments{This research was supported by the Israel Science Foundation (grant No. 1918/16).}

\bibliography{biblist.bib}
\end{document}